\documentstyle[preprint,cite,aps]{revtex}
\newcommand{\be}{\begin{equation}}
\newcommand{\ee}{\end{equation}}
\newcommand{\bea}{\begin{eqnarray}}
\newcommand{\eea}{\end{eqnarray}}
\newcommand{\ba}{\begin{array}}
\newcommand{\ea}{\end{array}}

\newcommand{\bb}{\bibitem}

\begin{document}
\draft

\title{\bf Reply on \lq\lq Lifshitz-point critical behaviour to $O(\epsilon_L^2)$''}
\author{Luiz C. de Albuquerque$^\ddagger$\footnote{
e-mail:claudio@fma.if.usp.br, lclaudio@fatecsp.br}
and Marcelo M. Leite$^\star$\footnote{e-mail:leite@fma.if.usp.br, leite@fis.ita.br}}
\address{}
\maketitle

\bigskip
\begin{center}
{\small

$\ddagger$ {\it Faculdade de Tecnologia de S\~ao Paulo -
FATEC/SP-CEETEPS-UNESP. Pra\c{c}a Fernando Prestes, 30, 01124-060.
S\~ao Paulo,SP, Brazil}
\bigskip

$\star$ {\it Departamento de F\'\i sica, Instituto
Tecnol\'ogico de Aeron\'autica, Centro T\'ecnico Aeroespacial,
12228-900,
S\~ao Jos\'e dos Campos, SP, Brazil}
}
\end{center}
\vspace{0.5cm}

\begin{abstract}
{\it We reply to a recent comment by H. W. Diehl and M. Shpot
(cond-mat/0106502) criticizing our paper J. Phys. A: Math. Gen. {\bf 34} 
(2001) L327-332. We show that the approximation we use for evaluating 
higher-loop integrals is consistent with homogeneity. A new renormalization 
group approach is presented in order to compare the two methods with 
high-precision numerical data concerning the uniaxial case. We stress that the 
isotropic behaviour cannot be obtained from the anisotropic one.}
\end{abstract}
\vspace{2cm}

\newpage

In a recent paper Diehl and Shpot \cite{DS1} (DS) criticized a
method proposed earlier in \cite{AL} to calculate the critical
exponents  $\nu_{L2}$, $\eta_{L2}$, and $\gamma_{L}$ at order
$O(\epsilon_L^2)$ for systems presenting an $m-$fold Lifshitz point.
Working entirely in momentum space we perform the calculations by using 
normalization conditions along with dimensional regularization. The 
symmetry point, used to define the renormalized theory, was chosen by setting the
external momenta scale along the quartic (competing) directions equal to zero,
while keeping the external momenta scale along the quadratic directions.
A detailed account of this approach was given in the second paper of 
reference \cite{AL} for the uniaxial ($m=1$) case.

In momentum space the most general solution to Feynman integrals
involving quadratic and quartic external momenta scale subspaces
perpendicular to each other is a difficult task, even at the one-loop level. 
Indeed, the one-loop integral $I_{2}$ contributing to the coupling constant 
can be performed exactly only if the external quartic momenta is set to zero.  
Keeping both external momenta scales, one can solve the integral over the 
quadratic momenta as a function of the external quadratic momenta, by choosing 
Schwinger parameters, for instance. The integral over the quartic momenta cannot be 
obtained as a function of the quartic external momenta in a closed form. Setting the 
quartic external momenta to zero simplifies the problem, since this integral will 
contribute with a simple factor to the remaining parametric integrals, which can 
be solved in a straightforward manner. Absorbing a convenient geometric 
angular factor, the result can be cast in a form which resembles the ordinary 
$\phi^4$ theory, with $\epsilon_{L}$ replacing $\epsilon$ and a slightly different 
coefficient for the regular term in $\epsilon_{L}$. The result is a homogeneous 
function of the quadratic external momenta scale.

The parametric integrals play a interesting role in our approximation. To see 
this, consider the simplest two-loop integral contributing to the two-point 
function, namely $I_{3}(p,k')$ given by

\begin{equation}
I_{3}(p,k') = \int \frac{d^{d-m}{q_{1}}d^{d-m}q_{2}d^{m}k_{1}d^{m}k_{2}}
{\left( q_{1}^{2} + (k_{1}^{2})^2 \right) 
\left( q_{2}^{2} + (k_{2}^{2})^2 \right) 
[(q_{1} + q_{2} + p)^{2} + \bigl((k_{1} + k_{2} + k')^{2}\bigr)^2]}\;\;.
\end{equation}

Setting $k'=0$, the integral can be evaluated as outlined in \cite{AL}. Before making 
our approximation, one can choose to integrate first either over the loop momenta 
$(q_{1}, k_{1})$ or over  $(q_{2}, k_{2})$. The loop integrals to be integrated 
first are referred to as the internal bubbles. By 
solving the integral over $q_{2}$ first, we obtain
\begin{eqnarray}
&& I_{3}(p, 0) = \frac{1}{2} S_{d-m} \Gamma(\frac{d-m}{2}) 
\int \frac {d^{d-m}q_1 d^{m}k_1}{q_{1}^{2} + (k_{1}^{2})^{2}} \nonumber \\
&& \int_{0}^{\infty} \int_{0}^{\infty} d \alpha_{1} d \alpha_{2} 
(\alpha_{1} + \alpha_{2})^{\frac{-(d-m)}{2}} 
exp(-\frac{\alpha_{1} \alpha_{2}}{\alpha_{1} + \alpha_{2}} (q_{1} + p)^{2}) 
\int d^{m}k_{2} e^{-\alpha_{1} (k_{2}^{2})^{2}} 
e^{-\alpha_{2} ((k_{1} + k_{2})^{2})^{2}}.
\end{eqnarray}

Now we can consider the approximation. In order to integrate over $k_{2}$, we have 
to expand the argument of the last exponential. This will produce a complicated function of 
$\alpha_{1}, \alpha_{2}, k_{1}$ and $k_{2}$. Unfortunately, this function has no 
elementary primitive. Considering the remaining terms as a damping factor 
to the integrand, the maximum of the integrand will be either at $k_{1}=0$ or 
at $k_{1} = -2k_{2}$. (The most general choice $k_{1}= -\alpha k_{2}$ yields a 
hypergeometric function.) The choice $k_{1} = -2k_{2}$ implies that $k_{1}$ varies 
in the internal bubble, but not arbitrarily. Its variation, however, is dominated by 
$k_{2}$ through this constraint, which eliminates the dependence on $k_{1}$ in the 
internal bubble. At these values of $k_{1}$, the integration over $k_{2}$ produces a simple 
factor to the parametric integral proportional to $(\alpha_{1} + \alpha_{2})^{-\frac{m}{4}}$. 
This allows one to perform the remaining parametric integrals in a simple way.
Thus, the constraint is designed to preserve the {\it form} of the parametric 
integrals. After realizing these integrals, they produce the factor 
$((q_{1} + p)^{2})^{-\frac{\epsilon_{L}}{2}}$. Note that the diagrams $I_{3}$ and $I_{5}$ 
contributing to the two-point function receive the factor $\frac{1}{2 - \frac{m}{4}}$ after 
integrating over the quadratic momenta in the external bubble.
This factor will not be present in the isotropic case, since there is no integration over 
quadratic momenta to be done in this case. The resulting 
solution to $I_{3}(p,0)$ is a homogeneous function of the external momenta $p$, not 
a generalized homogeneous function, given by:   
               
\begin{equation}
I_{3} = - (p^{2})^{1 - \epsilon_{L}} \frac{1}{8-m}\,\frac{1}{ \epsilon_{L}} 
\Biggl[1 + \biggl([i_{2}]_m + \frac{3}{4-\frac{m}{2}} + 1 \biggr)\epsilon_{L}\Biggr].
\end{equation}

The implementation of this constraint on higher-loop integrals proceeds 
analogously. The constraint turns all these integrals into homogeneous 
functions of the external quadratic momenta scale. One can then choose the 
symmetry point as $p^{2}= \kappa_{1}^{2}$, for example, in order to define the 
renormalized vertices via normalization conditions. The normalization 
constants $Z_{\phi}(\kappa_{1}), Z_{\phi^{2}}(\kappa_{1})$ and the beta function are defined in 
\cite{AL}, which give origin to the exponents $\nu_{L2}$ and $\eta_{L2}$ (under 
the momentum flow in the scale $\kappa_{1}$), along with all the scaling 
relations relative to exponents perpendicular to the competing axes 
\cite{rengro}. This follows in complete analogy to the usual $\phi^{4}$ 
theory describing the Ising model. As the constraint is based on a physical 
principle (homogeneity), we do not agree that the approximation is unacceptable. 

Actually, we agree that the speculation made in \cite{AL} ( ``... suggests that calculations 
performed in momentum space and coordinate space are inequivalent, as far as the Lifshitz point is 
concerned'') was unfortunate. (DS) work was the first one to extend the treatment done for the $m=2,6$ cases to general $m$ by making use of the scaling form of the free propagator in coordinate space 
representation. However, this does not give them support for their speculation that `` there is no way that 
AL's and our calculation can be both correct''. In the following we shall outline a new renormalization approach 
in momentum space of this problem using {\it two independent fixed points}, which is different from (DS) 
treatment using only one fixed point and will help the subsequent discussion.

So far, we have obtained half the solution to the problem, as we still have to devise a method to 
calculate the critical exponents along the competing axes. We follow a suggestion made by 
Wilson in the early seventies \cite{wilson} in order to obtain these exponents independently 
from the ones perpendicular to the competition axes. 
We can consider another independent set of 
normalization conditions defined at zero quadratic external 
momenta and nonvanishing {\it quartic} external momenta scale 
$\kappa_{2}$ \cite{rengro}. At the Lifshitz point, the free propagator have only quartic momenta 
along the competition axes. Thus, it is possible to perform a dimensional redefinition of the 
$m$-dimensional subspace by considering the associated quartic momenta to have half the dimension 
of a conventional momentum scale. As a consequence, the term in the bare Lagrangian 
which is proportional to the quartic momenta does not need to be multiplied by another dimensionful 
normalization constant ($\sigma_{0}$) in order to be meaningful on dimensional grounds. Under a flow 
in $\kappa_{2}$ at the corresponding fixed point, the normalization constants
$Z_{\phi}(\kappa_{2})$,$Z_{\phi^{2}}(\kappa_{2})$ lead to the critical exponents $\eta_{L4}, \nu_{L4}$ 
and new scaling laws along the competing axes, which are independent of the ones obtained in the subspace 
perpendicular to the competition directions. In this case, in order to evaluate loop integrals we use 
approximations which preserve homogeneity of the Feynman integrals in the external quartic 
momenta scale $\kappa_{2}$, such that scaling theory is fulfilled. Specifically, 
consider the one-loop integral $I_{2}(0,P)$:  
\be
I_2(0,P)=
\int \frac{d^{d-m}q d^m k}{\left( ((k + P)^{2})^{2}
+ q^{2}  \right) \left( (k^{2})^{2} + q^{2}  \right)}\;\;\;.
\ee

The simplest approximation for this integral which preserves
homogeneity in the quartic external momenta scale is 
$((k + P)^{2})^{2} = ((k)^{2})^{2} + ((P)^{2})^{2}$. Of course, more involved 
approximations can be developed which preserve homogeneity, but we concentrate on this 
one for the sake of simplicity. The result for
this integral is 
$I_{2}(0,P) = (((P)^{2})^{2})^{- \frac{\epsilon_{L}}{2}} 
\frac{1}{\epsilon_{L}} (1 + [i_{4}]_{m} \epsilon_{L})$, where 
the geometric angular factor 
$\frac{1}{2} \Gamma(\frac{m}{4}) \Gamma(2 - \frac{m}{4}) S_{d-m}
S_{m}$ has been absorbed in a redefinition of the coupling constant, 
and $[i_{4}]_{m} = \frac{1}{2}[1 + \psi(1) - \psi(2 - \frac{m}{4})]$. This result 
reflects the independent infrared divergence of this integral on the external 
momenta scale along the competition axes. The beta function for this case,  
$\beta(u) = -2 \epsilon_{L}(\frac{\partial lnu_{0}}{\partial u})_{\kappa_{2}}$, 
is different (and independent) from the one 
associated to critical exponents perpendicular to the competition
axes, even though both have the same fixed point at one-loop level. 
It can be easily checked that at the one-loop 
$\nu_{L4} = \frac{\nu_{L2}}{2}$. Thus, homogeneity is the guiding
principle for obtaining the solution to arbitrary loop integrals as a function of $\kappa_{2}$. 
The resulting scaling relations for exponents associated to correlations perpendicular to the 
competing axes are independent from the ones along the competition axes. 

The renormalization group just described can be adapted to treat the isotropic behavior 
$m=d$ close to 8. However, the isotropic case is intrinsically different from this 
renormalization group perspective. There is only one momenta scale $\kappa_{2}$ and just 
one set of normalization conditions. The beta function   
$\beta(u) = - \epsilon_{L}(\frac{\partial lnu_{0}}{\partial
  u})_{\kappa_{2}}$ is half the value of the one associated to the $\kappa_{2}$ 
characterizing the competing directions in the anisotropic case. They are different, 
since the coupling constant in both cases have different canonical dimension. Technically, 
the isotropic loop integrals do not receive contributions from the parametric integration 
over the quadratic momenta subspace, for they are absent in this case. That is why the 
results for the anisotropic behavior described in \cite{AL} cannot be extended to the 
isotropic one. The isotropic behavior has its own scaling relations, which 
are independent from the ones concerning the correlations along the competition axes for the 
anisotropic case \cite{rengro}.

We can now analyse the previous RG formalisms possessing only one independent momenta scale for the 
anisotropic case. The first modern treatment in terms of 1PI vertex parts was given by Mergulh\~ao and 
Carneiro \cite{mergulho}. There they set up the formalism 
in terms of normalization conditions in momentum space. They chose the symmetry point at nonvanishing 
quartic external momenta and zero quadratic external momenta as well as two conditions on the derivative of the two-point function at two independent external momenta scales. 
This reproduces the earlier scaling relations derived by Hornreich et al \cite{hornreich}. 
They went to coordinate space in order to calculate the exponents for the cases $m=2,6$. 
The novel feature of this approach is the 
introduction of an additional normalization constant $\sigma_{0}$, neeeded to obtain the exponents 
$\nu_{L4}, \eta_{L4}$, etc. In \cite{DS3}, (DS) followed this treatment entirely in coordinate space in order 
to extend the formalism to the general $m$-fold behavior. 
They introduced another normalization constant $\rho_{0}$ in order to treat the crossover and identified the 
critical exponents using the  renormalization group in coordinate space. 
The semianalytical coefficients in the $\epsilon_{L}$-expansion are integrals (generalized homogeneous functions) 
to be performed numerically in coordinate space. These numerical integrals only make sense if one splits the 
integration limits on the variable ${\bf v} = \sigma_{0} {\bf x_{\parallel}} x_{\perp}$ using the scaling and 
related functions in the coordinate space representation in the integrand up to the maximum value of ${|\bf v|}$ 
at ${|\bf v_{0}|} = 9.3$, and replacing the asymptotic values of these functions for greater values of 
${\bf v}$ \cite{DS3}. This numerical approximation is needed in order to obtain reasonable numerical values 
for the exponents $\eta_{L2}$ and $\eta_{L4}$ at $O(\epsilon_{L}^{2})$. Otherwise, the generalized homogeneous 
functions in the integrand are not suitable to describe properly the coefficients of the $\epsilon_{L}$-expansion.
They calculated $I_{2}$ and $I_{3}$ along these lines in reference \cite{DS3}.

After that, (DS) went to momentum space in order to calculate the two-loop integral 
$I_{4}({\bf Q},{\bf K})$ using dimensional regularization along with minimal subtraction 
\cite{DS2}. They used a mixed treatment, calculating some integrals in momentum
space, going to coordinate space whenever it was convenient (and 
vice-versa) and making use of a scaling function $\Phi$ (defined in 
equation (8) of \cite{DS2}). The integral $I_{2}$ is a subdiagram of $I_{4}$, depending on two external 
momenta scales as well. Nevertheless, they fixed the quartic external momenta scale to be zero and concluded 
that $I_{4}$ does not depend on it for general values of the quartic external momenta scale. 
Indeed, according to equation $(B.14)$ of \cite{DS2}
($\epsilon_L=4+\frac{m}{2}-d$)

\be
I_4(Q{\bf e}_\perp,{\bf K})=F^2_{m,\epsilon_L}\,\frac{Q^{-2\epsilon_L}}
{2\epsilon_L}\,\Bigl[\frac{1}{\epsilon_L}+J_u(m)+O(\epsilon_L)\Bigr].
\ee
This happens to be incomplete. The problem can be traced back to the
calculation of the one-loop integral:

\be
I_2(P_{\perp},K^\prime_{\parallel})=
\int \frac{d^{d-m}q d^m k}{\left( (k + K^{\prime})^{4}_{\parallel} 
+ (q + P)^{2}_{\perp} 
\right) \left( k^{4}_{\parallel} + q^{2}_\perp  \right)}\;\;\;.
\ee
They only computed this integral for vanishing  external momenta along
the quartic direction ($K^\prime_{\parallel}=0$). In this case, one has 

\be
I_2(P_{\perp},K^\prime_{\parallel}=0)=(P^2)^{-\frac{\epsilon_L}{2}}\,I_2({\bf
  e}_\perp).
\ee
However, by setting $P_{\perp}=0$, keeping $K^\prime_{\parallel}$
different from zero, we have from the discussion following Eq. (4)

\be
I_2(P_{\perp}=0,K^\prime {\bf e}_{\parallel})=(K^{\prime
  4})^{-\frac{\epsilon_L}{2}}\,I_2({\bf
  e}_\parallel),
\ee
We stress that $I_2({\bf e}_\parallel)$ and $I_2({\bf e}_\perp)$
are different functions in general. One can choose them to have the 
same leading singularities (the multiplicative factor is absorbed in
a redefinition of the coupling constant anyway), the difference
appearing in the regular terms in $\epsilon_L$. In fact, the complete
integral $I_2$ depends on these two momenta scales. 

We recall that in a proper minimal subtraction procedure,
{\it all} the external momenta should be kept {\it arbitrary} \cite{Amit}.
(DS) did not take into account this fact to proceed with the minimal 
subtraction, rather keeping only the quadratic external momenta
$Q{\bf e}_\perp$ in $I_4$ and $P_{\perp}$ in $I_2$,
and setting the quartic external momenta to zero in these integrals. 
They should show how the necessary cancellations of poles take place along the quartic 
subspace as well in order to have a satisfactory minimal subtraction scheme. In fact, 
the cancellations along the quadratic directions actually work when one fixes the 
quartic external momenta to zero. Even though this procedure is not complete, one can 
accept it as a new type of minimal subtraction to this problem.

For the anisotropic case, the (DS) method is based on one fixed point and the (almost exact) numerical integration for 
two-loop integrals which appear as the coefficients of the $\epsilon_{L}$-expansion using coordinate space 
representations whenever it is convenient. On the other hand, the method developed in reference \cite{rengro} (and discussed here)
in momentum space utilizes two fixed points. This new method states that there are four independent critical exponents (instead of three) 
with two independent set of scaling laws relating exponents along the quadratic and quartic directions in each 
subspace separately. We use an approximation, namely the 
constraint relating loop momenta in internal and external subdiagrams, which yields analytical results for higher-loop integrals.

A comparison of the two methods with numerical results for the exponents associated to perpendicular 
correlations to the competing axes (labeled with the subscript L2 after \cite{rengro}) is in order. 
For the $m=1, d=3, N=1$ case, (DS) found using MATHEMATICA: $\nu_{L2}=0.71, \gamma_{L2}=1.40$. This is 
consistent with the newest Monte Carlo simulations for $\gamma_{L2} = 1.36 \pm 0.03$ \cite{Henkel}, 
and compatible with an earlier Monte Carlo study ($1.40\pm 0.06$) \cite{Selke}. 
On the other hand, our approximation yielded  $\nu_{L2}=0.73$ and $\gamma_{L2}=1.45$
\cite{AL}. When using the new hyperscaling relation obtained in \cite{rengro} for the specific heat exponent, namely 
$2 - \alpha_{L2} = (d-\frac{m}{2})\nu_{L2}$ and replacing the value  $\nu_{L2}=0.73$, 
we obtain $\alpha_{L2} =  0.175$, whilst the most recent Monte Carlo calculation is 
$\alpha_{L2} = 0.18 \pm 0.02$ \cite{Henkel}. On the other hand, using the value obtained by (DS) $\nu_{L2}=0.71$ 
in the new hyperscaling relation, we find $\alpha_{L2}=0.225$. We can proceed and analyse the new scaling law 
obtained in \cite{rengro} for the magnetization exponent 
$ \beta_{L2} = \frac{1}{2} \nu_{L2} ((d-\frac{m}{2}) - 2 + \eta_{L2})$. Our calculation yields $\beta_{L2} = 0.198$, 
whereas the simulation result is $\beta_{L2} = 0.238 \pm 0.005$. Using (DS) results for $\nu_{L2}$ and 
$\eta_{L2}$ inside this new scaling relation one finds $\beta_{L2} = 0.192$. The high precision 
numerical values \cite{Henkel} are in very good agreement with the two-loop results using our approximation, which 
we believe cannot be said to be ``unacceptable'' at this point. The very similar values obtained for the exponents 
using either (DS) or our two-loop calculations confirms that momentum and coordinate space calculations should give the 
same results, since either our approximation or (DS) numerical approximation is responsible for a rather small 
deviation in the two results when compared to the above numerical output.

Finally, we emphasize the failure of (DS) method to treat the isotropic behavior. 
As was pointed out in \cite{AL} and explained in this work, $\nu_{L2}$ and $\eta_{L2}$ are 
not valid for the isotropic 
($m=8$) case. At the Lifshitz point, only the momenta scale along the
competing axes is meaningful for the isotropic case. Hence one has to
start from scratch using this momenta scale, which is incompatible
with our choice of normalization conditions. (DS) calculated $I_2$ and
$I_4$ along the components of quadratic external momenta only. But this
momenta scale makes no sense for the $m=8$ case at the Lifshitz point,
since they are not present any longer. In that case the fixed point
should be determined entirely as a function of the quartic external momenta
scale as shown in \cite{rengro}. There it was found that the isotropic behaviour cannot 
be obtained from the anisotropic behaviour. This is in contradiction to (DS) and we 
conclude that it is most likely the use of the momentum scale vanishing at $m=8$ that 
led (DS) to erroneous results. 

In conclusion, we have shown that our two-loop results do constitute a very good approximation for 
calculating critical indices. Our method proved to be very simple to give analytical expressions to the 
exponents. It is based on a renormalization group analysis consisting of two independent fixed points and is a natural alternative to the (DS) semianalytical 
approach based on only one fixed point. In view of the comparison with numerical values, we 
believe that both methods for the anisotropic cases deserve further investigation in 
order to unravel the fascinating issues concerning the Lifshitz critical behavior.

The authors would like to thank B. V. Carlson for a critical reading of the manuscript and support from FAPESP, 
grant numbers 00/03277-3(LCA) and 00/06572-6(MML).

\end{document}